\documentclass[aps, pra, reprint, longbibliography]{revtex4-1}

\usepackage{graphicx}
\usepackage{filecontents}
\usepackage{tabularx}
\usepackage{dcolumn}
\usepackage{bm}
\usepackage{amsmath}

\usepackage{natbib}
\usepackage{xr-hyper}
\usepackage{hyperref}
\usepackage{graphicx}
\usepackage{xcolor}    
\begin{document}

\preprint{APS/123-QED}

\title{Laser Melting and Loss of Interlayer Coherence of the Charge Density Wave Order in TiSe${_2}$ observed by time-resolved neutron diffraction}

\author{K. Dharmasiri}
\affiliation{Department of Physics, University of Virginia, Charlottesville, VA 22904, USA}
\author{S. S. Philip$^\bullet$}
\affiliation{Department of Physics, University of Virginia, Charlottesville, VA 22904, USA}
\author{S. A. Chen}
\affiliation{Neutron Scattering Division, Oak Ridge National Laboratory, Oak Ridge, TN 37831, USA}
\author{M. D. Frontzek}
\affiliation{Neutron Scattering Division, Oak Ridge National Laboratory, Oak Ridge, TN 37831, USA}
\author{Z. J. Morgan}
\affiliation{Neutron Scattering Division, Oak Ridge National Laboratory, Oak Ridge, TN 37831, USA}
\author{C. Hua}
\affiliation{Material Science and Technology Division, Oak Ridge National Laboratory, Oak Ridge, TN 37831, USA}
\author{D. Louca*}
\affiliation{Department of Physics, University of Virginia, Charlottesville, VA 22904, USA}

\date{\today}

\begin{abstract}
Time-resolved elastic neutron scattering combined with laser-induced heating was used to probe the evolution of the periodic lattice distortion (PLD) associated with charge-density-wave (CDW) order in 1T-TiSe$_{2}$. Upon pulsed laser heating, the crystal undergoes a phase transition, where the transient response of the average lattice is isotropic. In contrast, the PLD exhibits an anisotropic behavior. The superlattice intensity is rapidly suppressed in-plane and out-of-plane upon heating, consistent with melting of the ordered state, but recovers more quickly in-plane than along the c-axis as the sample cools below the CDW transition temperature. The out-of-plane component shows pronounced broadening due to reduction of the interlayer correlation length while the in-plane correlations recover. The disparity between the nearly isotropic lattice response and the anisotropic recovery of the superlattice demonstrates that the CDW dynamics cannot be described solely by the evolution of the average lattice temperature. Instead, the results indicate that interlayer phase coherence represents the slowest-relaxing component of the ordered state. These findings support a picture in which the CDW instability is established primarily within the 1T-TiSe${_2}$ layers, while three-dimensional long-range order depends on weaker coupling between layers. More broadly, they demonstrate how direction-resolved time-resolved neutron diffraction can disentangle thermal and correlation-driven contributions to CDW dynamics in low-dimensional materials.

\end{abstract}

\maketitle

Transition metal dichalcogenides (TMD) have the potential to outperform silicon in electronic and optoelectronic applications \cite{cheng2022, yu2023}. 1T-TiSe${_2}$, a van der Waals (vdW) type crystal (Fig. 1(a)) can fit this bill because it can be exfoliated to a monolayer and potentially find its way in solid-state cooling applications and phononic devices \cite{raya_moreno}. 1T-TiSe${_2}$ is also a prototype charge density wave (CDW) system and has been extensively studied to understand the mechanism of the CDW transition that occurs $\approx$ 200 K \cite{di1976electronic,wakabayashi1978phonons,akhanda2024}. There are two components to the CDW order parameter; one is associated with the formation of excitons, and the other with periodic lattice distortions (PLD) and electron-phonon coupling \cite{ou2023}. The PLD is manifested as a spatial modulation of atoms, with doubling of the unit cell in all three real-space directions (2a$\times$2b$\times$2c) and a new crystal symmetry \cite{porer2014,hedayat2019,wegner2}. 

An indirect exciton condensate arising from mixing states at $\Gamma$ and L points in the electronic band structure is one leading candidate mechanism for TiSe$_{2}$'s CDW, and because it forms at finite Q (not at Q = 0) it breaks translational symmetry and produces the observed density/lattice modulation. However, the extent to which the transition is purely electronically driven versus a joint electron-phonon instability remains an active experimental and theoretical question \cite{monney2009,van2010exciton,abbamonte,wegner2020evidence}. To disentangle the different components relevant to the CDW order, a number of techniques coupling ultrafast laser excitations and spectroscopy or scattering have been employed to capture the far-from-equilibrium dynamics \cite{zhang2020, duan2023,yang2017,caruso2022,lian,burian,weber2011,anharmonicity,raya_moreno,sun2018,ikeda2018, heinrich2023} of CDW melting, and measurements showed that the system can be driven across the CDW by breaking the excitons while subsequently destroying the PLD\cite{Vorobeva:10,Huber2024}. Most theoretical approaches\cite{singh2004, hsu2021, yin2024} predict that the PLD should collapse essentially instantaneously with the CDW. However, the two components respond differently to the optical excitations; even if excitons are quickly suppressed, the PLD was shown to survive in some excited state \cite{hedayat2019} at longer timescales. The electron and lattice degrees of freedom are typically described within a two-temperature model. The electron temperature is expected to rise sharply as electrons absorb energy and heat up quickly within 10-100 fs after which they exponentially decay \cite{perfetti2007,jia2025,maldonado2017,bennemann2004}. The lattice temperature, on the other hand, responds much more slowly as energy is transferred from the hot electrons via electron-phonon coupling\cite{mizukoshi2023}. Even though CDW melting is predominantly attributed to the breakdown of excitons\cite{lian,burian}, the role of phonons is arguably important as well \cite{weber2011,anharmonicity,raya_moreno}. 

Using a new technique where a transient thermoelastic pulse is generated across the surface of a sample by pulsed laser irradiation, the structural dynamics of the CDW are investigated in 1T-TiSe${_2}$ using time-resolved neutron diffraction. 
How the PLD reorganizes under rapid energy deposition, and the state to which 1T-TiSe${_2}$ settles as it is rapidly driven across the phase boundary are here explored using fast laser heating and time-resolved neutron diffraction. Fast laser heating accesses structural states not observed under equilibrium heating. It can be used to probe effects arising from nucleation, domain-wall motion, diffusion, or strain relaxation. Fast laser heating, combined with time-resolved neutron scattering, provides access to states that govern non-equilibrium phase evolution, without relying on coherent lattice dynamics. In this work, we observed that on heating, the PLD and lattice follow the same time-scales, but upon removal of the laser, the PLD reappears promptly in-plane as the temperature drops below the CDW transition but recovers more slowly out-of-plane. Any H- or L-direction asymmetry observed in the satellite peaks cannot be attributed to bulk heating/cooling profile. This asymmetry may point to an anisotropy of the CDW order parameter's dynamics. 
The in-plane CDW response is largely governed by transient heating and follows equilibrium behavior, whereas the out-of-plane correlations exhibit an additional nonequilibrium suppression manifested as significant peak broadening. This indicates that thermoelastic strain preferentially disrupts interlayer phase coherence and cannot be described solely by an elevated lattice temperature.


%
%
\begin{figure}[t!]
\includegraphics[width=3.4 in]{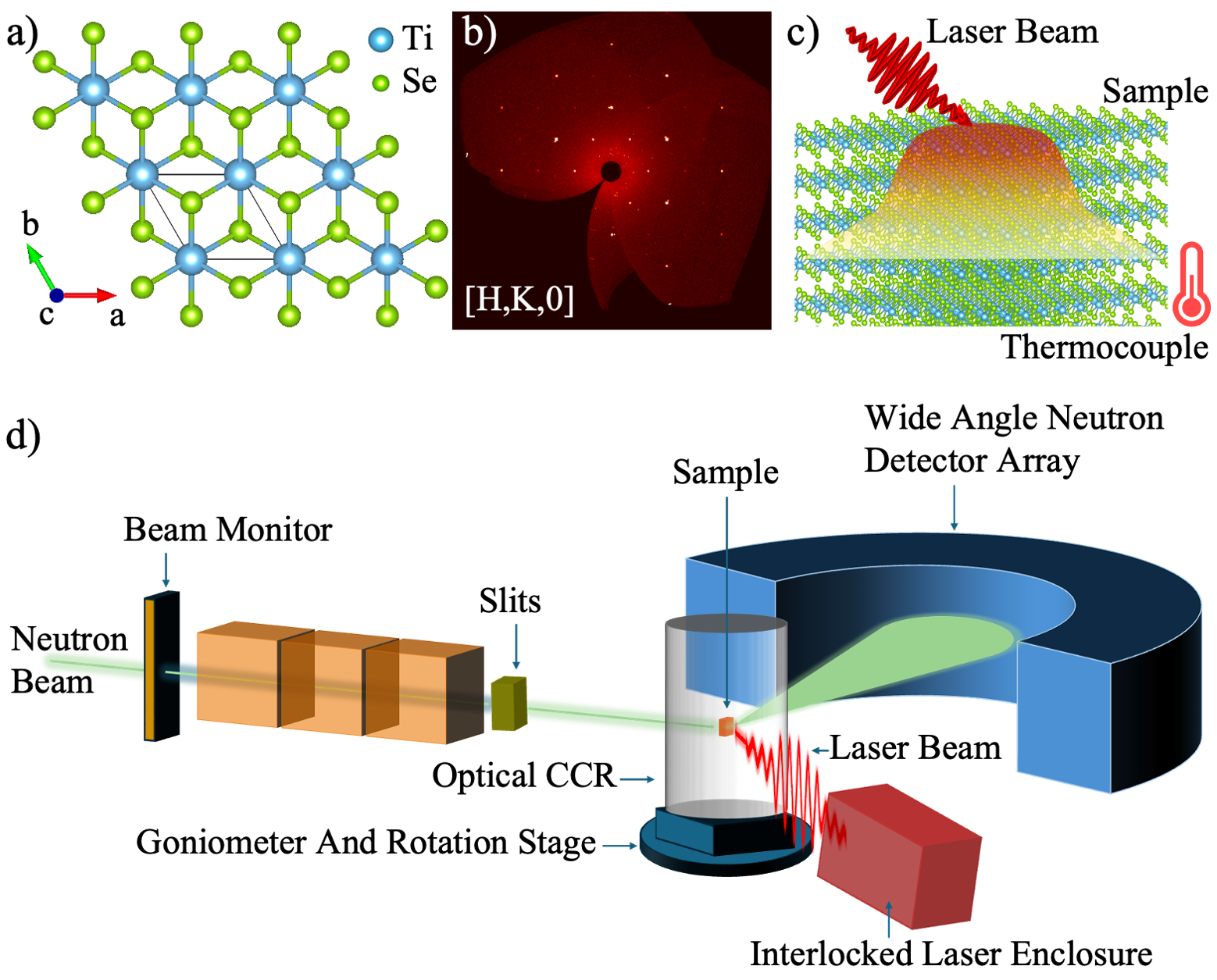}
\caption{(a) The crystal structure of TiSe$_{2}$. The atoms are arranged in a hexagonal lattice. (b) A plot of the diffraction pattern in the (hk0) plane, obtained from single crystal X-ray diffraction at room temperature. (c) An illustration of the laser heating used to induce a thermal gradient across the sample. A focused laser beam is directed at the sample, creating distinct hot and cold regions. The thermocouple is in contact with the base of the sample. (d) A schematic of the neutron scattering experimental setup. The neutron beam enters from the left and passes through a beam monitor and a series of slits that shape and collimate the beam. The sample is mounted on a goniometer and rotating stage. It is housed within an optical closed-cycle refrigerator (CCR). A wide-angle neutron detector array is positioned to capture scattered neutrons, and an interlocked laser enclosure directs an additional laser beam at the sample.}
\label{fig1}
\end{figure}

\begin{figure}[t!]
\includegraphics[width=3.4 in]{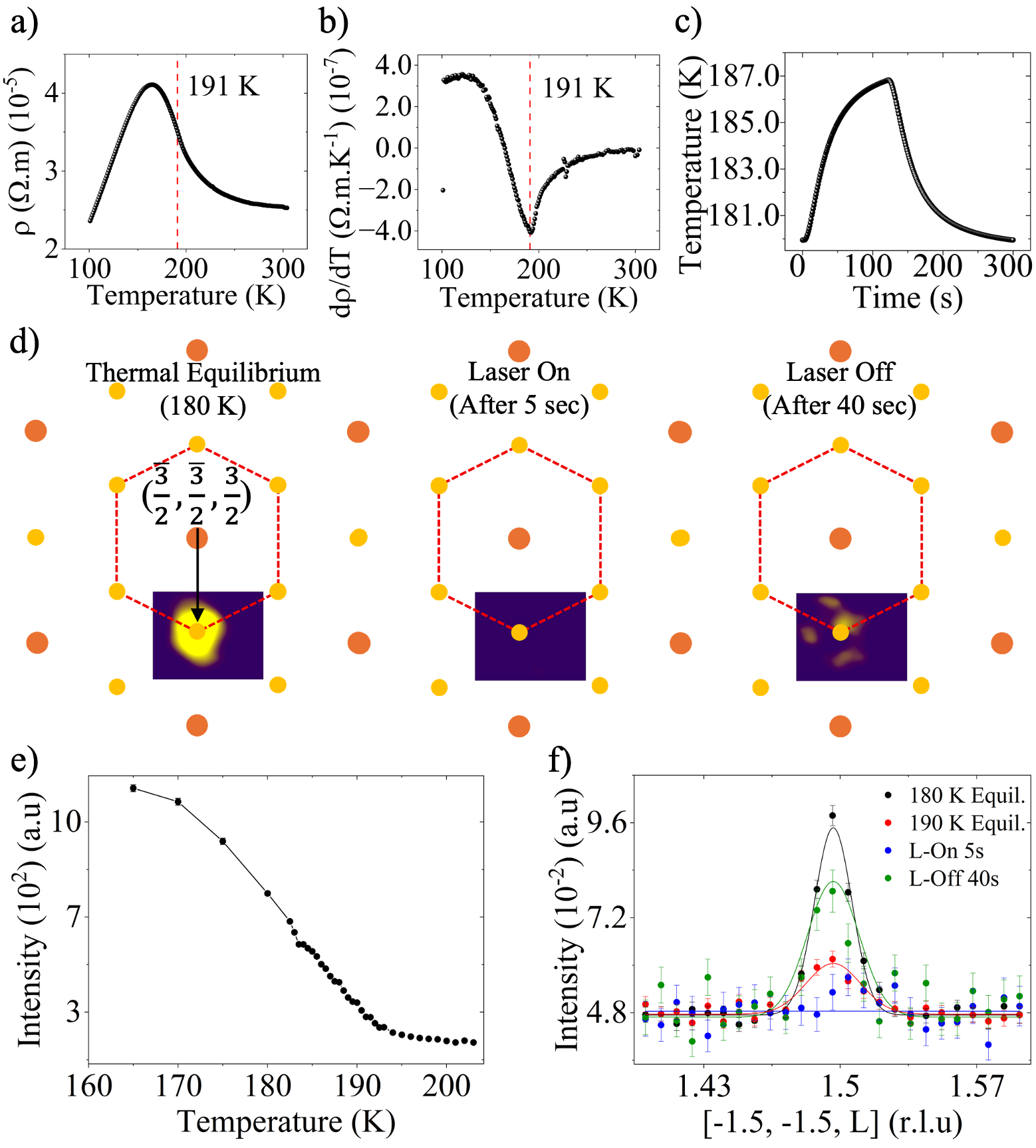}
\caption{(a) A plot of electrical resistivity ($\rho$) as a function of temperature (T) from 100 to 300 K and $\rho$ in units of $\Omega\cdot$m. The red dashed line at approximately 191 K marks the CDW transition. (b) A plot of the derivative of $\rho$ with respect to temperature, $\frac{d\rho}{dT}$, is shown. A dip is observed around 191 K, marked by a red dashed line. (c) A plot of the temperature as measured by the thermocouple. The first 120 seconds is with the laser on and the next 180 seconds is with the laser off. (d) Plots of the PLD under different conditions. The first, labeled "Thermal Equilibrium" at 180 K, shows the $(\frac{\bar3}{2}$, $\frac{\bar3}{2}$, $\frac{3}{2})$ superlattice across a hexagonal grid. The second, labeled "Laser On" after 5 seconds, reveals a noticeable absence in intensity at the ($\frac{\bar3}{2}, \frac{\bar3}{2}, \frac{3}{2}$) point. The third, labeled "Laser Off" after 40 seconds, shows the intensity returning. (e) A plot of the integrated intensity at the superlattice as a function of temperature at equilibrium. (f) A comparison of the intensity of 180 and 190 K equilibrium data, the laser on for 5s, and the laser off for 40s.}
\label{fig2}
\end{figure}

The single crystal of TiSe$_2$ was grown using chemical vapor transport \cite{HQ_graphene}, and single-crystal diffraction measurements confirmed its high quality (Fig. 1(b)). Shown in Fig.~\ref{fig2}(a) is the temperature-dependent resistivity, $\rho(T)$, of TiSe${_2}$. The derivative of $\rho(T)$ shows a sharp drop at 191 K, marked by a red dashed line in the figure, that marks the CDW transition temperature. For the neutron experiment at WAND$^2$ at Oak Ridge National Laboratory (ORNL), the crystal was cut to an area of less than 1 cm$^2$. Measurements were carried out both above and below the CDW transition using a neutron wavelength of $\lambda = 1.486$ Å. To drive the system across the phase boundary, we employed a laser heating setup as shown in Fig.~\ref{fig1}(d) and also in Fig. S1 in the Supplement \cite{supplement}. A pulsed laser (Coherent Flare NX, Class 4) operating at 515 nm ($\sim$2.4 eV) with a repetition frequency up to 2000 Hz, a pulse width of ~5 ns, and a pulse energy of 300 µJ was used. The laser fluence at the sample surface was 1.53 mJ/cm$^{2}$, with approximately 75\% absorbed by the sample \cite{bealhughesliang}. The laser beam diameter at the sample position was 5 mm. The laser was alternately turned on for 120 s and off for 180 s. Fig. \ref{fig2}(c) shows the rise and fall of the cold-finger temperature during the 300 s cycle. The thermocouple temperature reading, determined by its relative distance from the sample and the thermal conductance at the interface between the thermocouple and the sample holder, as discussed in the Supplement \cite{supplement}, may not be accurate. The volume-average sample temperature was estimated from the shift of the fundamental Bragg peak as discussed below. The on/off sequence was repeated 22 times. Event-based data collection mode was used to time-bin the data into one-second intervals, making this the first measurement of its kind to combine a pulsed laser with neutron diffraction. Additional details of the experimental setup are provided in the Supplement \cite{supplement}.

When the experimental timescale is on the order of seconds, the system evolves into a long‑time thermal regime characterized by local equilibrium between the sample and its environment. Diffraction measurements with seconds‑level temporal resolution therefore primarily capture slow structural relaxation and phase evolution \cite{jia2025}. In this long-timescales regime, a single temperature heat diffusion model, described by C(T)$\frac{\delta{T}}{\delta{t}}$=$\nabla\cdot(\kappa(T)\nabla{T}$)+Q(t,r), is appropriate since electronic and lattice degrees of freedom remain in thermal equilibrium at all times relevant to the experiment \cite{karani2015, shayduk2020}. Here, C(T) is the heat capacity, $\kappa(T)$ is the thermal conductivity and Q(t,r) is the laser heat source. Upon laser excitation, energy is deposited within the near‑surface region, producing a transient temperature gradient that drives diffusive heat transport into the bulk. In semiconducting TiSe$_{2}$, both electrons and phonons contribute to thermal transport \cite{mizukoshi2023}. Although TiSe$_{2}$ exhibits strong electron-phonon coupling  \cite{sadasivam2017}, electron–phonon equilibration occurs on ultrafast timescales far shorter than the experimental resolution. As a result, the single‑temperature model provides a reasonable and self‑consistent description of heat flow and structural evolution in the long‑time limit probed by diffraction.

Since the laser spot size is comparable to the in-plane cross-sectional dimensions of the crystal, heat transport can be approximated as one-dimensional along the c-axis. Longitudinal heat flow depends on the sample thickness; in this case, the crystal thickness was $L = 0.5$ mm. The characteristic timescale of heat propagation can be estimated as $\Delta t \sim L^2/D$, where $D = \kappa(T)/C_v(T)$ is the thermal diffusivity, giving $\Delta t \approx 200$ ms. The laser pulse interval was 0.5 ms, much shorter than this thermal relaxation time, implying that heat accumulates in the sample. A steady-state temperature profile was established within the first 100 ms after the laser was turned on (detailed modeling is provided in the Supplement \cite{supplement}). Cooling of the sample can be treated within the lumped-capacitance heat-transfer approximation, with a timescale $\tau = L C_v(T)/G$, where $G$ is the thermal interface conductance between the sample and its holder. Across the surface of the sample, $\kappa(T)$ is at least 4 times larger that along the c-axis. Given the laser spot size and sample size, $\Delta t \approx 1-5$ s.

Shown in Fig.~\ref{fig2}(d) are momentum-space resolved images under three conditions: (1) thermal equilibrium at 180 K (pre-laser), (2) after 5 seconds of laser-on, and (3) after 40 seconds with the laser-off. These images, mapped over a hexagonal reciprocal lattice, show how laser heating first suppresses the PLD, and when the laser heating source is turned off, how it recovers. Shown in Fig. 2(e) is the integrated superlattice intensity of the $(\frac{\bar3}{2}$,$\frac{\bar3}{2}$,$\frac{3}{2})$ peak at equilibrium (pre-laser), consistent with the transition observed in the transport data of Figs.~\ref{fig2}(a) and \ref{fig2}(b). In Fig. \ref{fig2}(f), the intensity of the superlattice from data collected at 180 K (black dots) (pre-laser) is compared to the intensities with the laser-on after 5s (blue dots), with the laser-off after 40s (green dots), and to the 190 K equilibrium data (pre-laser) (red dots). The intensity drops substantially after laser heating, almost to zero within 5s. The intensity is less than the 190 K equilibrium data which shows that in spite of the temperature reading of the thermocouple at the cold-finger ($\approx187$ K as seem in Fig. \ref{fig2}(c)), the actual sample temperature is above the transition. An estimate of the actual sample temperature is discussed further below.

\begin{figure*}[t]
\begin{center}
\includegraphics[width=0.95\textwidth]{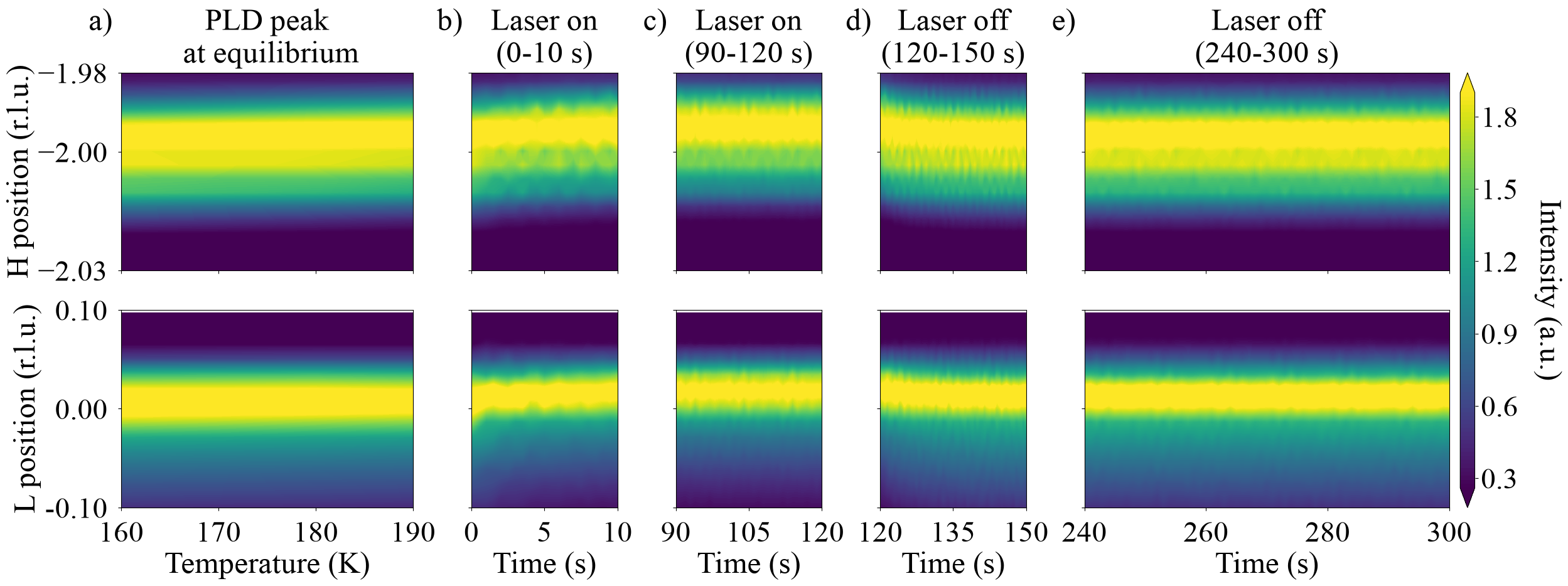}
\end{center}
\caption{The time evolution of the (${\bar{2},\bar{2},0}$) Bragg peak. The top panels are plots of the peak profile along the H position (r.l.u.), while the bottom panels are for the L direction (r.l.u.). (a) The Bragg peak profile at equilibrium as a function of temperature (160 to 190 K). b–e) Time-resolved evolution of the peak where the laser is turned on for 2 minutes and then off for 3 minutes: b) the initial response when the laser is turned on (0–10 s); c) the driven steady-state while the laser remains on (90–120 s); d) the thermal relaxation after the laser is turned off (120–150 s); and e) the late-stage recovery with the laser off (240–300 s). The color bar indicates diffraction intensity in arbitrary units (a.u.).}
\label{fig3}
\end{figure*}

We next characterize the response of the fundamental (${\bar{2}\bar{2}0}$) Bragg reflection to the laser excitation, in order to establish a baseline for purely lattice/thermal effects uncomplicated by the CDW order parameter (Fig.~\ref{fig3}). Panel (a) shows the equilibrium peak profile across 160-190 K, exhibiting an expected thermal response of the reflection. Upon laser excitation ((b), 0-10 s), the peak changes positions slightly along both H and L, consistent with the temperature rising as the heat from the laser propagates through the illuminated volume. This persists into the driven steady state ((c), 90-120 s) and recovers when the laser is switched off ((d), 120-150 s), as the peak relaxes back to its equilibrium profile in both H and L by late times ((e), 240-300 s). Critically, the Bragg peak shows no directionally selective change in position or intensity between H and L: full recovery is observed on both axes. This rules out a trivial anisotropic Debye-Waller (thermal-disorder), and confirms that the fundamental lattice periodicity is unaffected within the timescale and the fluence used throughout this experiment.

\begin{figure}[t]
\includegraphics[width=3.4 in]{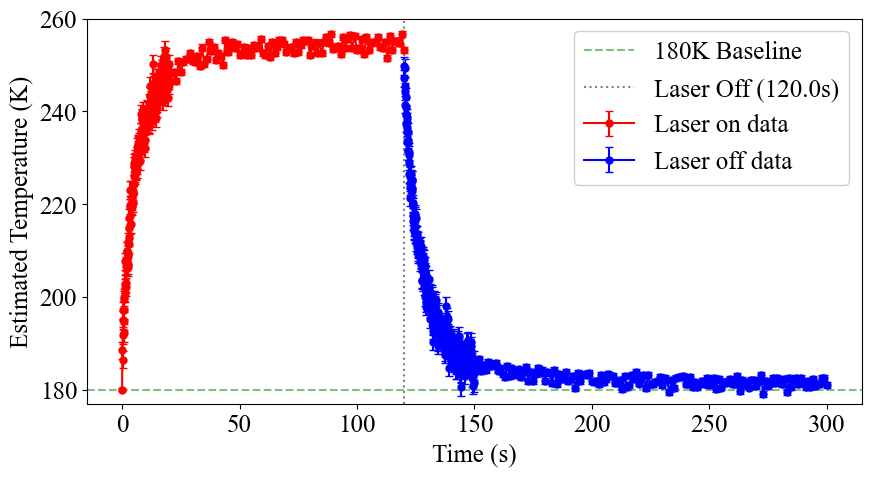}
\caption{A plot of the sample temperature as a function of time determined from the peak shift of (${\bar{2},\bar{2},0}$) Bragg peak along H. During the heating phase from 0 to 120 seconds, a rapid temperature increase from the 180 K baseline to a steady state near 255 K is observed. The vertical grey dotted line at 120 seconds marks the transition when the laser is turned off. The blue curve forllows the subsequent thermal relaxation phase from 120 to 300 seconds, as the temperature drops back to the 180 K baseline. To capture the rapid initial dynamics, the first 20 seconds of the laser-on phase and the first 30 seconds of the laser-off phase are time-binned at 0.1 s; all remaining data points are time-binned at 1 s.}
\label{fig4}
\end{figure}

To quantify the transient sample temperature independently of the CDW order parameter itself, we extracted an effective temperature from the thermal shift of the (${\bar{2}\bar{2}0}$) fundamental Bragg peak position along H, using the equilibrium temperature data shown in Fig.~\ref{fig3}(a). Because this reflection is a fundamental lattice peak, its position tracks the ordinary thermal expansion of the crystal and provides a purely lattice-based thermometer, uncomplicated by any CDW-specific structural response. Fig.~\ref{fig4} shows the resulting temperature trace over the full laser on/off cycle, averaged over all cycles. From the 180 K baseline, the sample heats rapidly upon laser illumination, reaching a quasi-steady-state temperature of approximately 255 K within the first several seconds and remains near this value for the duration of the 120 s laser-on window. When the laser is switched off at t=120 s, the temperature relaxes back toward the 180 K baseline; the bulk of this relaxation occurs within approximately 20-30 s, with the sample asymptotically approaching the baseline over the following minutes. Both the heating and cooling transitions were measured with finer time binning (0.1 s) over the first 20-30 s of each phase to resolve the fast initial dynamics, while the remainder of each phase was binned at 1 s given the comparatively slow subsequent relaxation. This temperature trace confirms that the laser excitation drives the sample from below T$_{CDW}$ to a steady-state temperature above it. Second, because the temperature dependence was derived entirely from the fundamental lattice peak, it establishes an isotropic, purely thermal baseline against which direction-selective behavior of the CDW satellite reflections are compared.


\begin{figure*}[t]
\begin{center}
\includegraphics[width=0.95\textwidth]{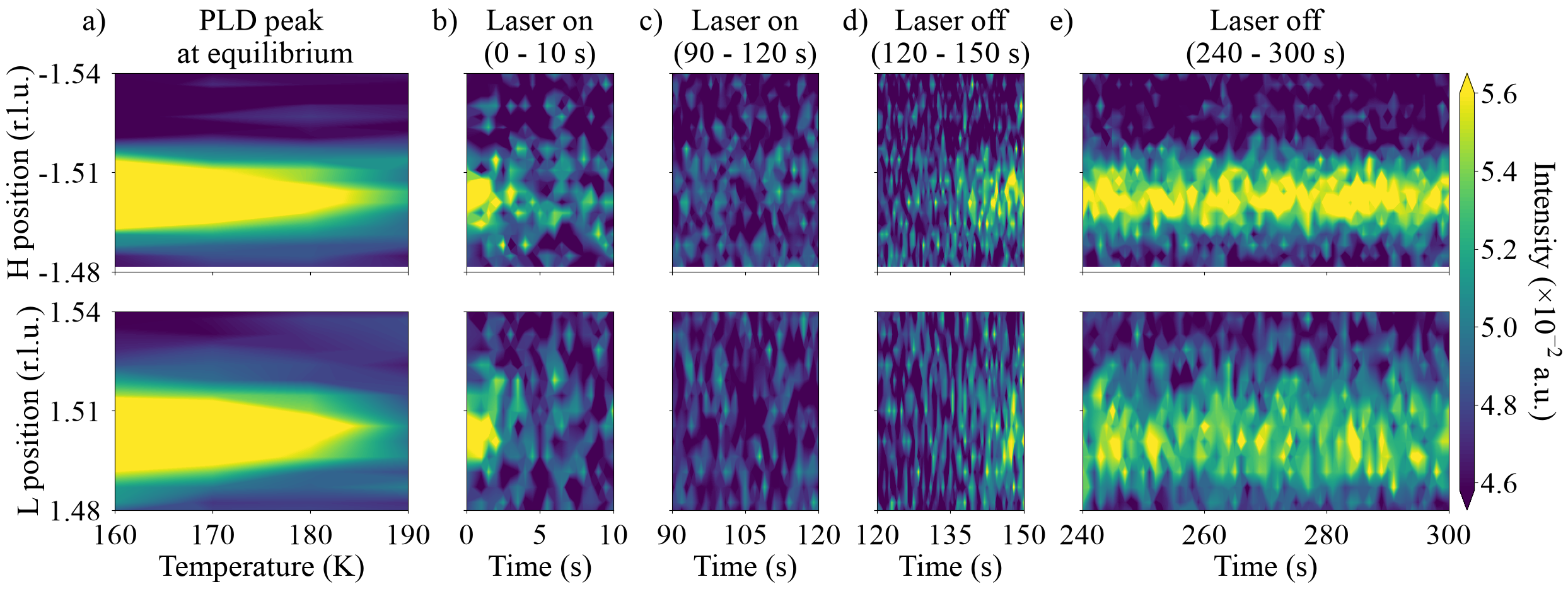}
\end{center}
\caption{The time evolution of the $(\frac{\bar3}{2}$,$\frac{\bar3}{2}$,$\frac{3}{2})$ PLD peak. The top panels are plots of the peak profile along the H position (r.l.u.), while the bottom panels are plots along L (r.l.u.). (a) The PLD peak profile at equilibrium as a function of temperature (160 K to 190 K). (b)–(e) Time-resolved evolution of the peak where the laser is turned on for 2 minutes and then off for 3 minutes: (b) the initial response when the laser is turned on (0–10 s) shows the PLD peak disappearing within 6 seconds; (c) the driven steady-state while the laser remains on (90–120 s); (d) the thermal relaxation after the laser is turned off (120–150 s), with the PLD peak beginning to reappear 15 seconds after the laser is switched off; and (e) the late-stage recovery with the laser off (240–300 s). The color bar indicates diffraction intensity in arbitrary units (a.u.).}
\label{fig5}
\end{figure*}

Having established that the fundamental lattice reflection shows no directionally selective response to the laser excitation (Fig.~\ref{fig3}) and using the associated effective-temperature trace as the thermal reference (Fig.~\ref{fig4}), we now turn to the PLD satellite reflection at $(\frac{\bar3}{2}$$\frac{\bar3}{2}$$\frac{3}{2})$ which directly tracks the CDW order parameter. At equilibrium, the PLD peak grows as the sample is cooled through T$_{CDW}$, reaching full intensity by 160 K (Fig.~\ref{fig5}(a)). Upon laser excitation, the peak intensity collapses rapidly: within the first several seconds of illumination the PLD reflection has essentially vanished into the background. This is clearly seen both along H and L (Fig.~\ref{fig5}(b)). This suppressed state persists throughout the laser-on window, with no resolvable PLD intensity recovering during the driven steady-state phase (Fig.~\ref{fig5}(c)), indicating that the order parameter remains fully melted for as long as the sample is held near the ~255 K steady-state temperature.

When the laser is switched off, the PLD peak does not recover instantaneously. Diffuse, low-level intensity begins to reappear only after approximately 20 s (Fig.~\ref{fig5}(d)), which compares to the lattice thermal relaxation observed in the Bragg peak, where the bulk of the temperature drop occurs within a comparable time window. By late times (Fig.~\ref{fig5}(e), 240-300 s), the PLD peak has substantially recovered in both H and L, though the recovered peak appears broader and less sharply defined than the pre-pump equilibrium profile in panel (a), suggesting that full re-establishment of long-range CDW order may require longer than the 180 s window captured here. 

\begin{figure}[t]
\includegraphics[width=3.4 in]{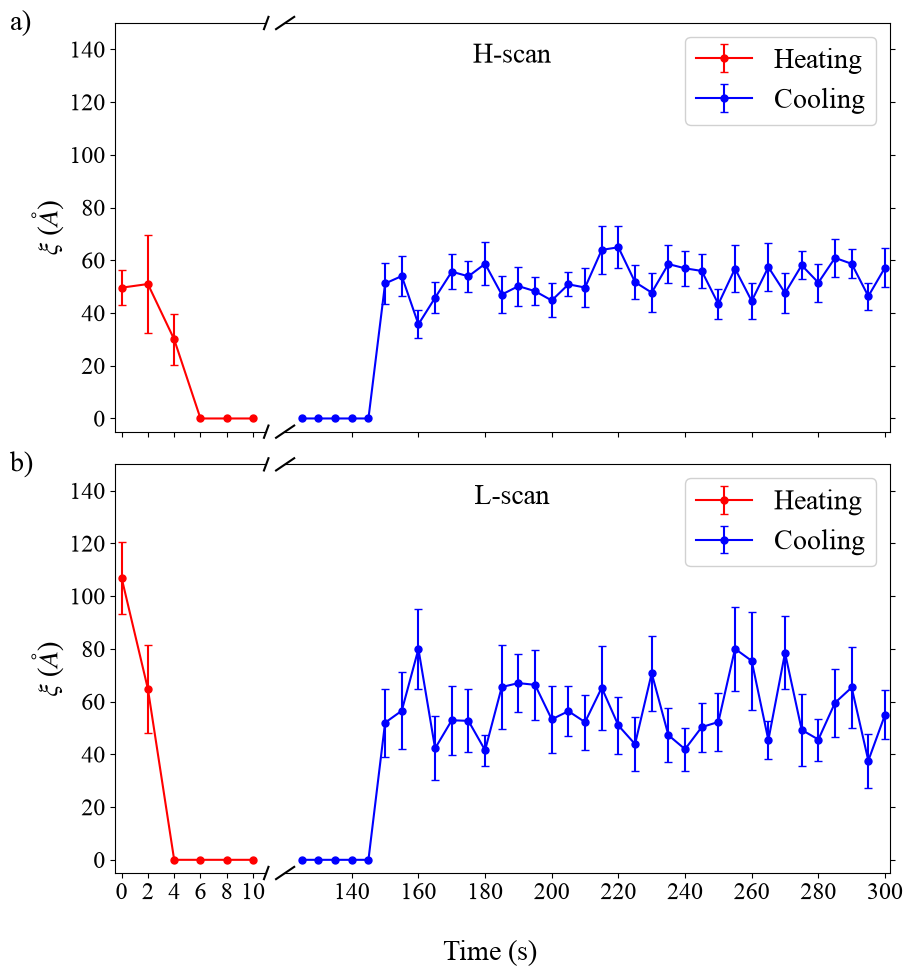}
\caption{Time evolution of the CDW peak correlation length ($\xi$) along the (a) H and (b) L directions. $\xi$ is determined from the full width at half-maximum (FWHM) of Lorentzian fits to the corresponding diffraction peaks. During the heating phase, laser heating suppresses the CDW order, driving $\xi$ to zero within approximately 5 s. During the cooling phase, the CDW peak and a finite correlation length reappear after the laser is turned off.}
\label{fig6}
\end{figure}

The fast lattice cooling (shown in Fig. 4) matches the reappearance of CDW order (Fig. 5) indicating that the CDW order parameter tracks the instantaneous lattice temperature. However, it can be seen in Fig.~\ref{fig5}(e), that in the L-direction, the intensity is compromised relative to the H-direction. This suggests that interlayer phase coherence is slower to recover than in-plane CDW order. Re-establishing long-range order along H (in-plane) may simply require realigning phase within a layer, whereas along L, it requires re-locking the relative stacking phase between layers. This process may be slower and does not complete in the time scale of the experiment. By 240-300 s, in-plane order has substantially recovered while interlayer coherence is still catching up. This difference in the response between H and L is also seen in the correlation length, $\xi$, obtained from fitting the peak in the H and L direction. Note that $\xi$ stays the same on heating on cooling in the H-direction, but drops to about half in the L-direction (Fig.~\ref{fig6}).

Fig.~\ref{fig6} (a) and (b) show how the $\xi$ changes during the heating and cooling cycles along H and L directions respectively. During the initial 120 sec of heating, the $\xi$ collapses to 0 withihn 5 seconds, dropping from approximately 50 \text{\AA}  along the H direction and 100 \text{\AA} along the L direction. During the cooling phase, the CDW peak reemerges within  30 seconds and the $\xi$ fluctuates around 50 \text{\AA} in both directions. 

 A wide range of laser‑induced phenomena have been observed in low‑dimensional quantum materials. For example, in 1T-TaS${_2}$, ultrafast laser excitation has been shown to melt the Mott‑insulating CDW state and induce a metastable metallic phase \cite{vaskivskyi2015}, while in VO${_2}$, optical pumping can drive an insulator‑to‑metal transition by disrupting the Peierls distortion \cite{wegkamp2015}. In TiSe${_2}$, previous studies have investigated lattice recovery dynamics and energy dissipation mechanisms on timescales relevant for potential device applications, such as light‑induced phase control in correlated electron systems \cite{monney2016}. In this work, time‑dependent measurements combining neutron diffraction and laser heating capture the slow suppression and reformation of the PLD on sub-seconds timescales. By combining time-resolved diffraction from a fundamental Bragg reflection with time-resolved measurements of the $(\frac{\bar3}{2}$$\frac{\bar3}{2}$$\frac{3}{2})$ CDW/PLD satellite, we separate the purely thermal/lattice response of the crystal from the response of the CDW order parameter. The Bragg-peak temperature dependence shows that the lattice heats to a quasi-steady-state temperature well above T$_{CDW}$ within seconds of laser illumination and relaxes back toward baseline on a comparable, few-tens-of-seconds timescale once the laser is switched off. The CDW satellite intensity tracks this same overall timescale, collapsing promptly upon heating and beginning to recover promptly upon cooling, indicating that, to first approximation, the CDW order parameter follows the lattice temperature rather than exhibiting a strongly decoupled or slower intrinsic relaxation. However, the recovery is not isotropic: at a late-time delay after the laser is switched off, the satellite intensity along H (in-plane) recovers to a substantially higher level than along L (out-of-plane/interlayer), even though both directions sample the same reflection and are therefore free of any structure-factor or geometric bias that could otherwise account for the difference. This directional asymmetry cannot be explained by the lattice temperature dependence alone, since the underlying temperature trace extracted from a fundamental Bragg peak shows no H/L asymmetry and is isotropic.

We interpret these results as evidence that the translational degrees of freedom associated with CDW order are strongly anisotropic. The in-plane component, linked to the quasi-two-dimensional electronic structure, recovers on a timescale consistent with the evolution of the lattice temperature, whereas interlayer phase coherence, governed by comparatively weak coupling between adjacent layers, relaxes more slowly and remains incompletely restored over the same interval. This behavior indicates that the establishment of local CDW order and the recovery of three-dimensional stacking coherence are distinct processes. Consistent with the quasi-two-dimensional character of the CDW state, robust intralayer order is rapidly recovered following cooling, while long-range interlayer phase coherence remains comparatively fragile. The recovery of superlattice intensity demonstrates re-establishment of the local CDW amplitude, whereas the persistent broadening of the 00l reflections reveals incomplete recovery of interlayer phase coherence. Together, these observations distinguish amplitude recovery from phase-ordering dynamics and identify the interlayer stacking phase as a soft collective degree of freedom whose evolution is not determined solely by the instantaneous lattice temperature.`


The authors would like to thank G.-W. Chern, L. Zhigilei, S. Nagler and A. Tenant for helpful discussions, Jie Xing for setting up the closed cycle refrigerator and Diane Dickie for help with the single crystal X-ray diffractometer. This work was made possible by the support from the National Science Foundation, Grant No. 2219493. C.H. was supported by the U.S. Department of Energy (DOE), Office of Science, Office of Basic Energy Science (BES), Materials Sciences and Engineering Division. A portion of this research used resources at the High Flux Isotope Reactor, a DOE Office of Science User Facility operated by the Oak Ridge National Laboratory. The beamtime was allocated to WAND${^2}$ on proposal numbers IPTS-30312.1 and 31675.1. The data can be obtained from the following repository \cite{github_data}.

$\ast$To whom correspondence should be addressed: $louca@virginia.edu$.

$^\bullet$New address: SLAC National Accelerator Laboratory, Stanford University, Menlo Park, CA 94025.

\bibliography{bibliography}

\clearpage

\end{document}


\title{Supplementary Material: Laser melting and loss of interlayer coherence of the charge density wave order in TiSe${_2}$ observed by time-resolved neutron diffraction}

\author{K. Dharmasiri}
\affiliation{Department of Physics, University of Virginia, Charlottesville, VA 22904, USA}
\author{S. S. Philip$^\bullet$}
\affiliation{Department of Physics, University of Virginia, Charlottesville, VA 22904, USA}
\author{S. A. Chen}
\affiliation{Neutron Scattering Division, Oak Ridge National Laboratory, Oak Ridge, TN 37831, USA}
\author{M. D. Frontzek}
\affiliation{Neutron Scattering Division, Oak Ridge National Laboratory, Oak Ridge, TN 37831, USA}
\author{Z. J. Morgan}
\affiliation{Neutron Scattering Division, Oak Ridge National Laboratory, Oak Ridge, TN 37831, USA}
\author{C. Hua}
\affiliation{Material Science and Technology Division, Oak Ridge National Laboratory, Oak Ridge, TN 37831, USA}
\author{D. Louca*}
\affiliation{Department of Physics, University of Virginia, Charlottesville, VA 22904, USA}

\maketitle

\beginsupplement

\section{Experimental configuration}
\begin{figure}[H]
\centering
\includegraphics[width=4 in]{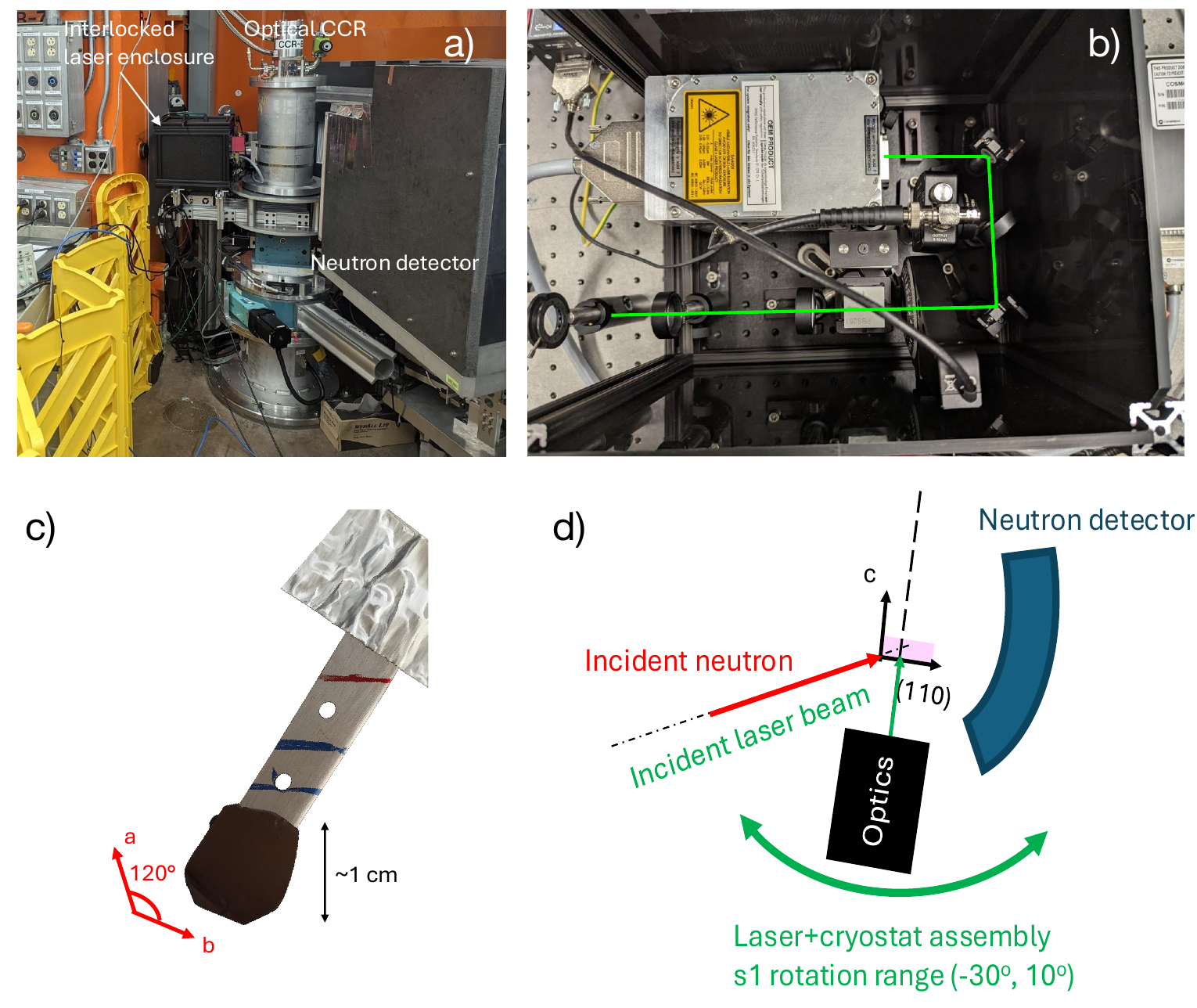}
\caption{(a) Laser setup integrated into WAND${_2}$ with an optical CCR. (b) The optical layout (Right) is much simpler compared to HYSPEC laser assembly due to the limited space at WAND${_2}$. (c) TiSe$_2$ crystal used in this measurement. (d) Schematics of neutron scattering configuration with respect to the laser assembly.}
\label{fig:ExpSetup}
\end{figure}

At WAND${_2}$ (HB-2C, HFIR), an optical setup on a 12''x10'' Thorlabs aluminum breadboard is mounted on 2'' rails (80/20 Inc) which are attached to the sample rotation stage as shown in Figure~\ref{fig:ExpSetup}(a).  A pulsed laser (Coherent Flare NX, Class 4) that operates at 515 nm ($\sim$ 2.4 eV) with a repetition frequency up to 2000 Hz, a pulse width of around 5 ns, and a pulse energy of 300 $\mu J$ is used. The laser can be externally triggered by TTL signals. The optical layout is shown in Figure~\ref{fig:ExpSetup}b. Right after the laser head is a motorized power control unit made of a rotating halfwave plate and cubic polarizing beam splitter (PBS). This unit enables a remote control of the laser power delivered to the sample. After the PBS, the light is horizontally polarized. Two 1'' lenses are used immediately after the PBS. The 1:2 ratio of focus lengths is used such that laser beam is enlarged to a diameter of 5 mm at the sample position. 

Figure \ref{fig:ExpSetup}(c) shows the TiSe$_2$ crystal used in this study. The in-plane cross sectional area ($ab$ plane) of the sample is about 1 x 1 cm and the thickness (along $c$ direction) is about 500 $\mu$m. Since the superlattice peaks associated with the charge density wave (CDW) state are observed in the [H H L] plane, the crystal must be rotated about the vertical axis parallel to the [H -H 0].  Considering the geometry of the laser enclosure and restricted space at WAND${_2}$, the accessible rotation angle, $s1$, of the laser/cryostat assembly is restricted to $-35^\circ < s1 < +10^\circ$, avoiding interference of the laser enclosure with the upstream incident neutron beam on the negative angle side and with either the downstream neutron beam or the scattered beam on the positive angle side (shown in Figure~\ref{fig:ExpSetup}d). This $s1$ range allows us to fully cover the superlattice peak at $(\frac{\bar3}{2}$,$\frac{\bar3}{2}$,$\frac{3}{2})$ and the Bragg peak at (${\bar{2},\bar{2},0}$). 

The experiments were conducted using an optical closed-cycle refrigerator capable of reaching a base temperature of 20 K. Equilibrium diffraction measurements were first performed at various temperatures above and below the phase transition with a neutron wavelength of $\lambda$ = 1.486 \AA, enabling full characterization of the superlattice peak intensity and line shape as a function of temperature. As shown in Fig. 2e in the main text, the superlattice peak intensity vanishes above 200 K and fully develops below 160 K. To drive structural changes, we employed a pulsed-laser heating scheme while maintaining the sample environment at 180 K. The laser was applied for 120 s at a repetition rate of $f_L = 2000$ Hz, delivering an energy fluence of approximately 1.53 mJ/cm$^{2}$. The measured superlattice peak intensity demonstrates that this fluence is sufficient to completely melt the CDW state. 

An analog-to-digital readout card (ADCROC) is utilized in this experiment to acquire data when the laser is on and off separately. The data reduction is performed using Mantid software \cite{Arnold}. For each data set, time binning, based on the time stamps of neutron counts,is carried out to filter the two-dimensional diffraction data into time bins, averaging over many data sets. In this way, time-dependent diffraction measurements can be generated even in a continuous neutron source.

To interpret the observed melting and recondensation dynamics of the PLD and their relation to the system’s thermal response, we present in the next section a thermal transport model describing the crystal’s temperature evolution under the applied laser-heating conditions. 

\section{Thermal transport}

The temperature response of the system is described by the heat diffusion equation:
\begin{equation}\label{eq:heat}
C_v(T)\frac{\partial T}{\partial t} =\kappa(T) \frac{\partial^2 T}{\partial z^2} + Q(z,t),
\end{equation}
where $C_v(T)$ and $\kappa(T)$ denote the volumetric heat capacity and thermal conductivity of TiSe$_2$, respectively, and $Q(\mathbf{r},t)$ represents the volumetric heat generation term induced by the pulsed laser. Since $C_v(T)$ and $k(T)$ vary only weakly in the temperature range between 180 K and 200 K, they are treated as constants with values $C_v = 1.6 \times 10^6$ J/K and $\kappa = 2$ W/m-K. Because the laser spot size is comparable to the in-plane cross-sectional dimension of the sample, heat transport can be approximated as one-dimensional along the $z$-direction or along the $c$-direction. The boundary conditions are given as $\partial T/\partial z |_{z=0} = 0$ and $\partial T/\partial z |_{z=L} = G[T_\infty-T(z=L)]$, where $L$ is the thickness of the crystal, $G$ is the interface conductance between the crystal and the sample holder, and $T_\infty$ is the environment temperature. Since TiSe$_2$ is a semimetal, the optical penetration depth of a 515 nm laser beam, $\delta$, is on the order of tens of nanometers. The volumetric heat generation term can be simplified to be $Q(\mathbf{r},t) = Q(t)exp(-z/\delta)$, where $\delta \approx 10$ nm. Solving Eq.~(\ref{eq:heat}) under a periodic drive  will allow us to investigate the thermal response of the system induced by the pulsed laser. 

\begin{figure}
\includegraphics[width=4 in]{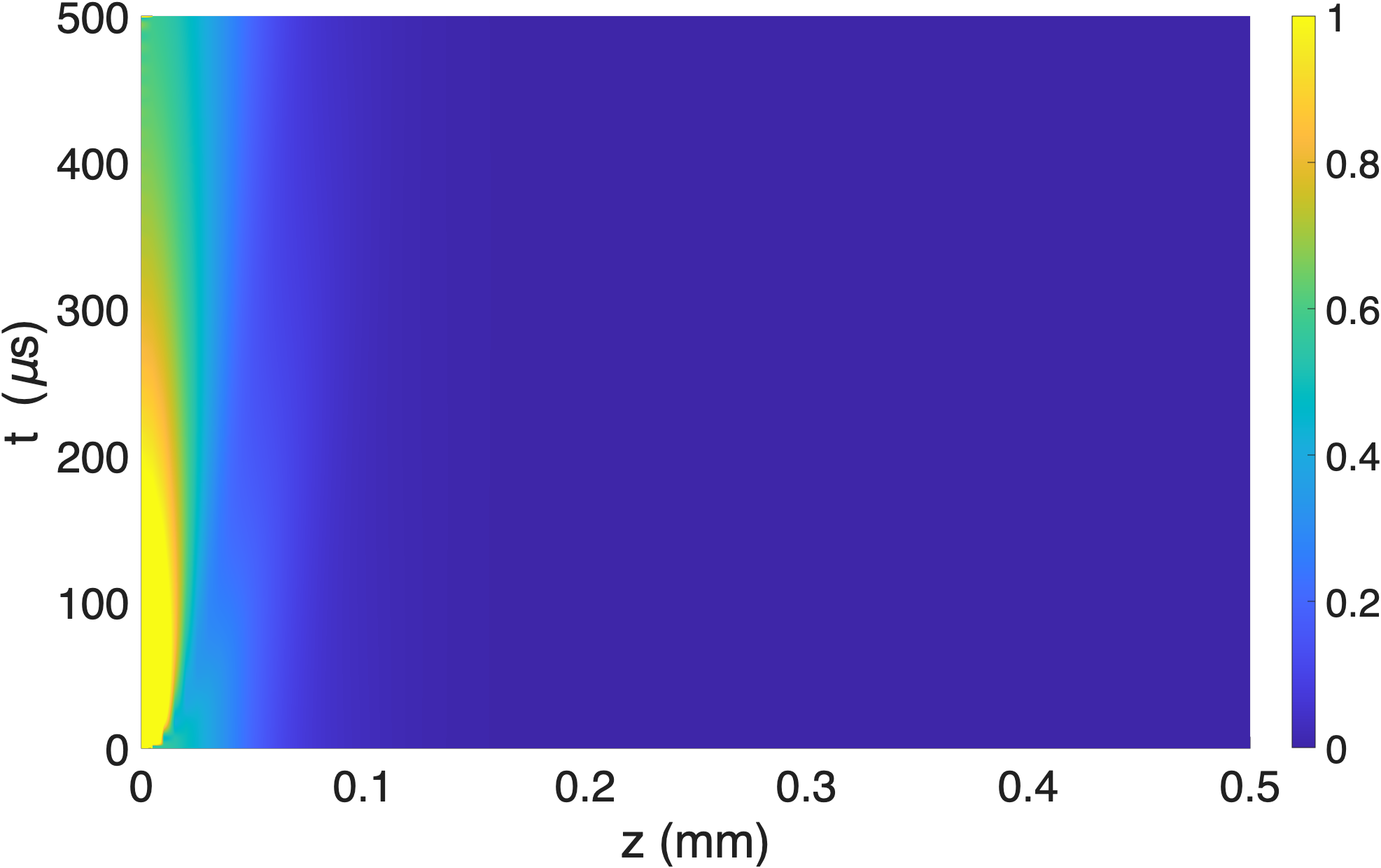}
\caption{Temperature response to a single laser pulse with a pulse width of 5 ns}
\label{fig:SinglePulseResponse}
\end{figure}

\begin{figure}
\includegraphics[width=4 in]{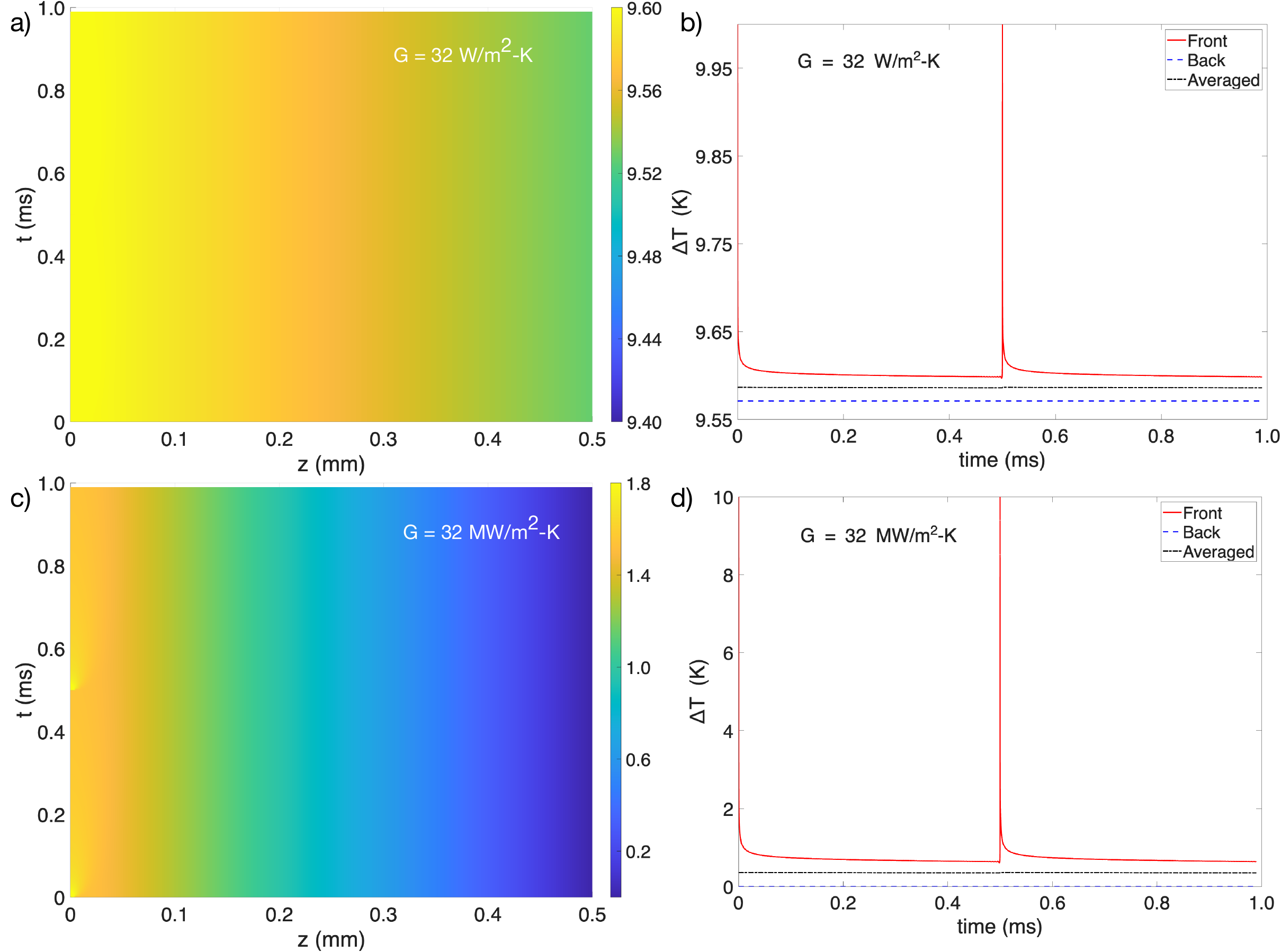}
\caption{Steady-state temperature profiles of TiSe$_2$ under a periodic laser excitation for interface thermal conductance values of (a) 32 W/m$^2$-K and (c) 32 MW/m$^2$-K. (b) For small conductance ($G \ll \sqrt{f_L C_v \kappa}$), heat is trapped in the crystal and the overall temperature gradient remains small. (d) For large conductance ($G \gg \sqrt{f_L C_v \kappa}$), heat is efficiently removed, resulting in a pronounced temperature gradient along the $z$-direction.}
\label{fig:MultiplePulseResponse}
\end{figure}

Figure \ref{fig:SinglePulseResponse} shows the temporal and spatial temperature response of the system to a single laser pulse, described by $Q(t) = Q_0 \exp(-t/\tau)$ $(t > 0)$, with $\tau \sim 5$ ns. When the system is continuously subjected to laser pulses at a repetition rate of 2000 Hz, heat accumulates at the front surface and gradually propagates toward the back surface. The characteristic timescale of this heat propagation can be estimated as $\delta t \sim L^2/D$, where $D = \kappa/C_v$ is the thermal diffusivity, giving $\delta t \approx 30$ ms in this case. This indicates that, after approximately 100 ms, the system reaches a steady state in which the temperature profile in TiSe$_2$ becomes periodic between successive cycles. The fully developed spatiotemporal temperature distribution depends strongly on the interface thermal conductance between the crystal and the sample holder. Figure \ref{fig:MultiplePulseResponse} illustrates the steady-state temperature profiles for two cases, where the interface conductance differs by several orders of magnitude. If the interface thermal conductance is small ($G \ll \sqrt{f_LC_v\kappa}$), the leaking of heat through interface is slow and the overall temperature gradient in the crystal is small (shown in Fig.~\ref{fig:MultiplePulseResponse}b). On the other hand, if the interface thermal conductance  is big ($G \gg \sqrt{f_LC_v\kappa}$), heat can be efficiently removed from the system, a large temperature gradient along the z-direction is then developed  (shown in Figure~\ref{fig:MultiplePulseResponse}d). 

Regardless of the interface thermal conductance value, the system’s thermal response to laser heating reaches a steady state within tens of milliseconds. In contrast, and as shown in Figure 4 in the main text, the intensity of the superlattice peak associated with the CDW state requires several seconds to fully vanish. This disparity indicates that the structural phase transition occurs on a timescale much slower than the thermal equilibration, suggesting the presence of dynamical processes during rapid annealing. 

\begin{figure}[H]
\centering
\includegraphics[width=6 in]{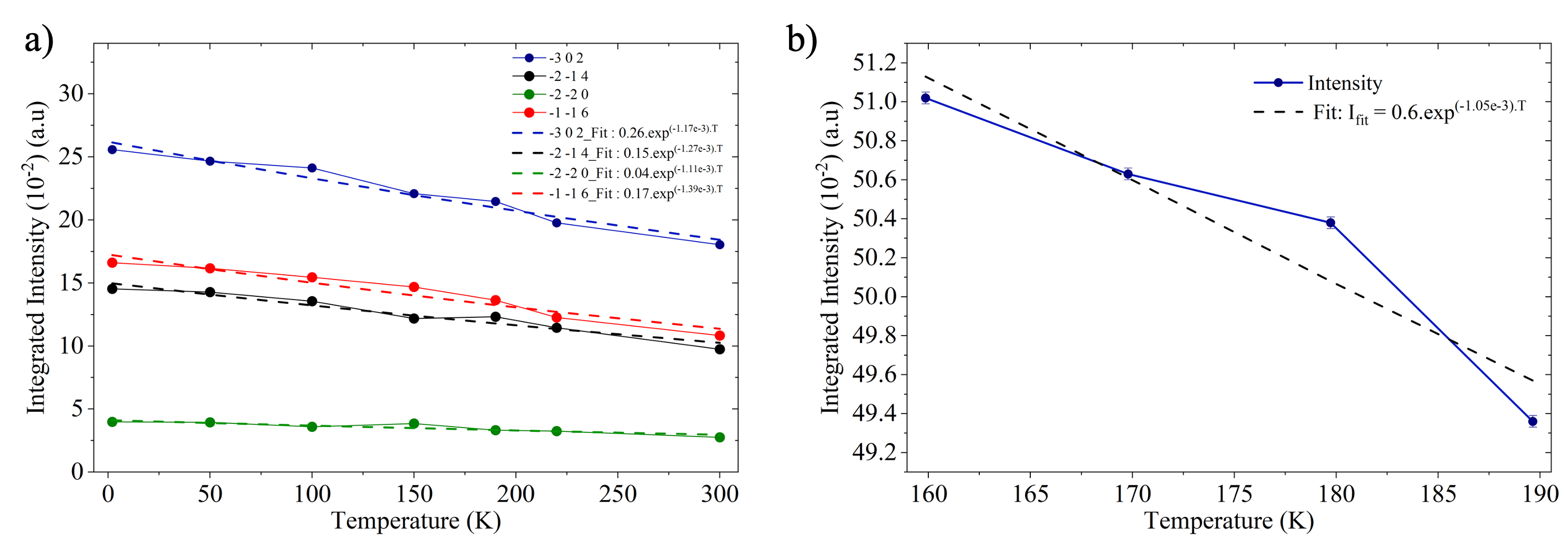}
\caption{Temperature dependence of Bragg peak intensities. (a) Exponential fits demonstrating the evolution of Bragg peak intensities as a function of temperature, obtained from NOMAD data (IPTS: 10595). (b) Corresponding exponential fit for the temperature evolution of the (${\bar{2},\bar{2},0}$) Bragg peak intensity obtained using the WAND$^2$ instrument.}
\label{fig:temp_est}
\end{figure}

Figure~\ref{fig:temp_est}(a) illustrates the temperature-dependent evolution of Bragg peaks within the Q range of 6 - 7 \AA${^{-1}}$, measured at various equilibrium temperatures using the NOMAD diffractometer for the TiSe$_2$ sample. Exponential fits to these data yield a decay constant of approximately $-1.2 \times 10^{-3}$ K$^{-1}$. Figure~\ref{fig:temp_est}(b) displays the analogous temperature evolution for the (${\bar{2},\bar{2},0}$) Bragg peak.
Analysis of both datasets reveals a consistent decay constant for the (${\bar{2},\bar{2},0}$)  reflection. Based on the observed fluctuations in the Bragg peak intensity, this decay constant ($-1.05 \times 10^{-3}$ K$^{-1}$) was used to estimate the in-situ temperature of the sample during the time-resolved laser measurements as shown in Figure 3 in the main text. 

\section{Superlattice peak evolution over time}

\begin{figure}[H]
\centering
\includegraphics[width=5 in]{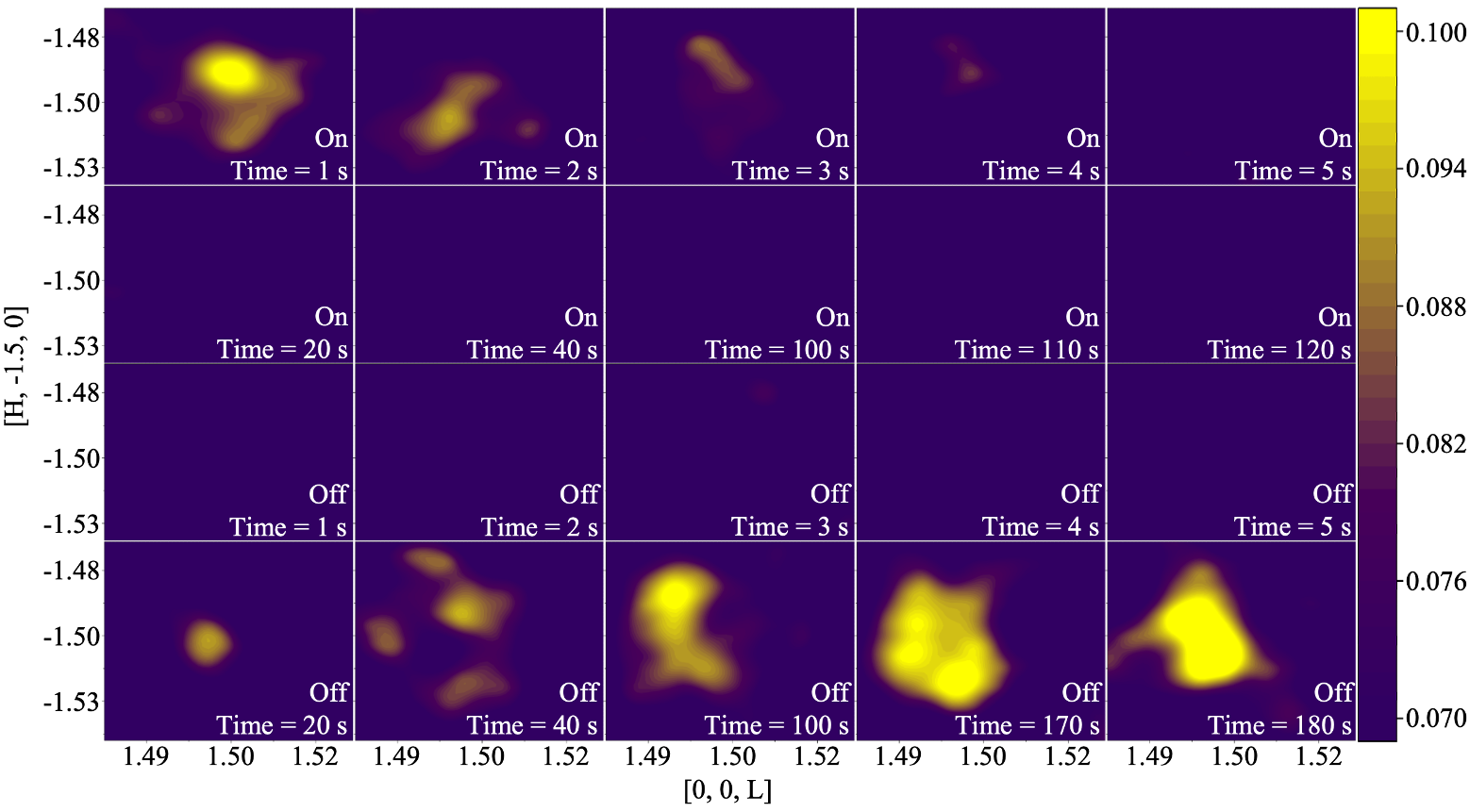}
\caption{\label{diagram}The plots show how the superlattice peak at $\frac{\bar3}{2},\frac{\bar3}{2},\frac{3}{2}$ evolves over time under two experimental conditions. The "On" and "Off" labels indicate when the laser is turned on for 2 minutes and then off for 3 minutes. The inset is a comparison of the integrated intensity of the on and off data at 60 seconds.}
\label{fig:supp_2D_data}
\end{figure}

Shown in Fig.~\ref{fig:supp_2D_data} are time‑resolved momentum‑space maps of the superlattice peak at  $(\frac{\bar3}{2}$,$\frac{\bar3}{2}$,$\frac{3}{2})$ under laser excitation and its subsequent recovery after the laser is turned off. The panels are obtained from event‑binned neutron diffraction data with temporal resolution on the order of a  second, and represent snapshots of the system at selected time intervals during the “Laser On” and “Laser Off” phases. The first row shows the “Laser On” condition, with representative time points ranging from 1 to 5 seconds after the onset of laser illumination (additional panels are provided in the Supplement). The diffraction intensity is displayed using a color scale ranging from blue (low intensity) to yellow (high intensity). Upon laser exposure, the superlattice peak intensity is strongly suppressed within the first few seconds accessible to our time resolution, indicating a rapid loss of long‑range PLD order. The second row shows the system during the “Laser Off” phase. After the laser is turned off, intensity gradually reappears at the $(\frac{\bar3}{2}$,$\frac{\bar3}{2}$,$\frac{3}{2})$ superlattice position, signaling a recovery toward the pre‑excited PLD phase. Over the full 180‑second duration of the laser‑off period, the diffraction maps exhibit a progressive increase in intensity and a re‑concentration of the signal in momentum space, consistent with thermal dissipation and the slow structural recovery of the original periodic lattice distortion.

\begin{figure}[H]
\centering
\includegraphics[width=5 in]{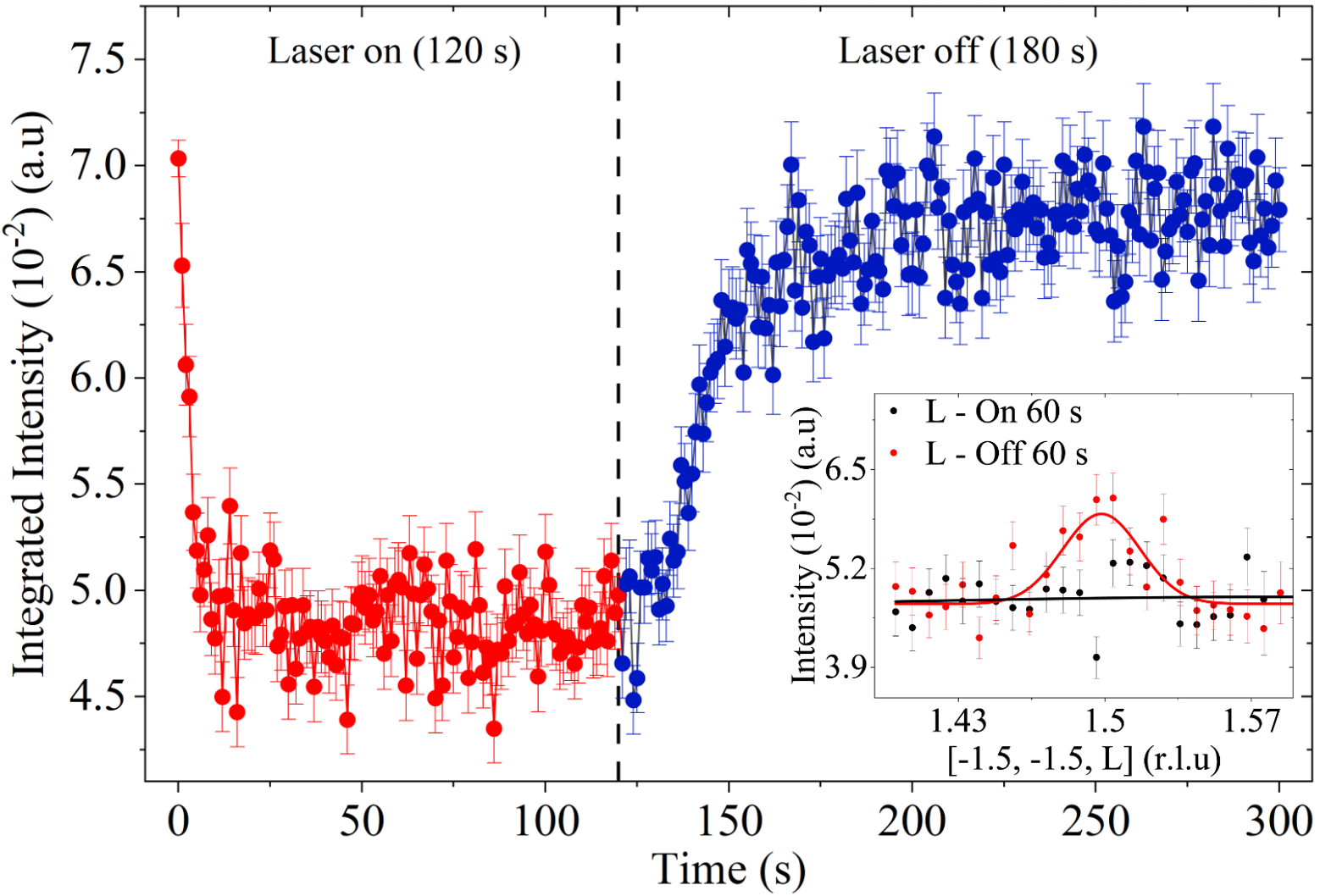}
\caption{\label{fig:Figure5} A plot of the integrated intensity and how it changes over a 300-second period, with a focus on the effect of a laser being turned on and off. The red colored region (0–120 s) labeled "Laser On" is a plot of the $(\frac{\bar3}{2}$, $\frac{\bar3}{2}$, $\frac{3}{2})$ integrated intensity over a 2 minute period. The blue colored region (120–300 s) labeled "Laser Off" is the integrated intensity of the same peak but with the laser off for 3 minutes. The data were integrated from -1.54 to -1.46 along the H direction, from -1.534 to -1.474 along the K direction and from 1.48 to 1.53 along the L direction. The error bars is the propagated error from the neutron measurement. During laser On (0–120 s), the intensity drops dramatically within seconds. During laser Off (120–300 s), the intensity of the $(\frac{\bar3}{2}$, $\frac{\bar3}{2}$, $\frac{3}{2})$ is mostly recovered within 30 to 40 seconds. The inset is a comparison of the intensity at 60s laser on and off.}
\label{fig:Comparison}
\end{figure}

Figure~\ref{fig:Comparison} is a graph depicting the integrated intensity as a function of time over a 300-second interval, confirming the reversible nature of the PLD. The data is divided into "Laser On” (0–120 s), highlighted in red, and “Laser Off” (120–300 s), highlighted in green. During the “Laser On” phase, the intensity of the $\frac{\bar3}{2}, \frac{\bar3}{2}, \frac{3}{2}$ state decreases after 1 second and remains low at the baseline after 5 seconds, indicating the suppression or destruction of PLD. After the 120th second, the “Laser Off” phase is initiated, and the intensity gradually increases, suggesting the reappearance of the CDW peak. The recovery of PLD takes approximately 30-40 seconds. The inset provides a cut across the PLD at 60 seconds, comparing the “Laser On” and “Laser Off” states. 

\begin{figure}[H]
\centering
\includegraphics[width=4.6 in]{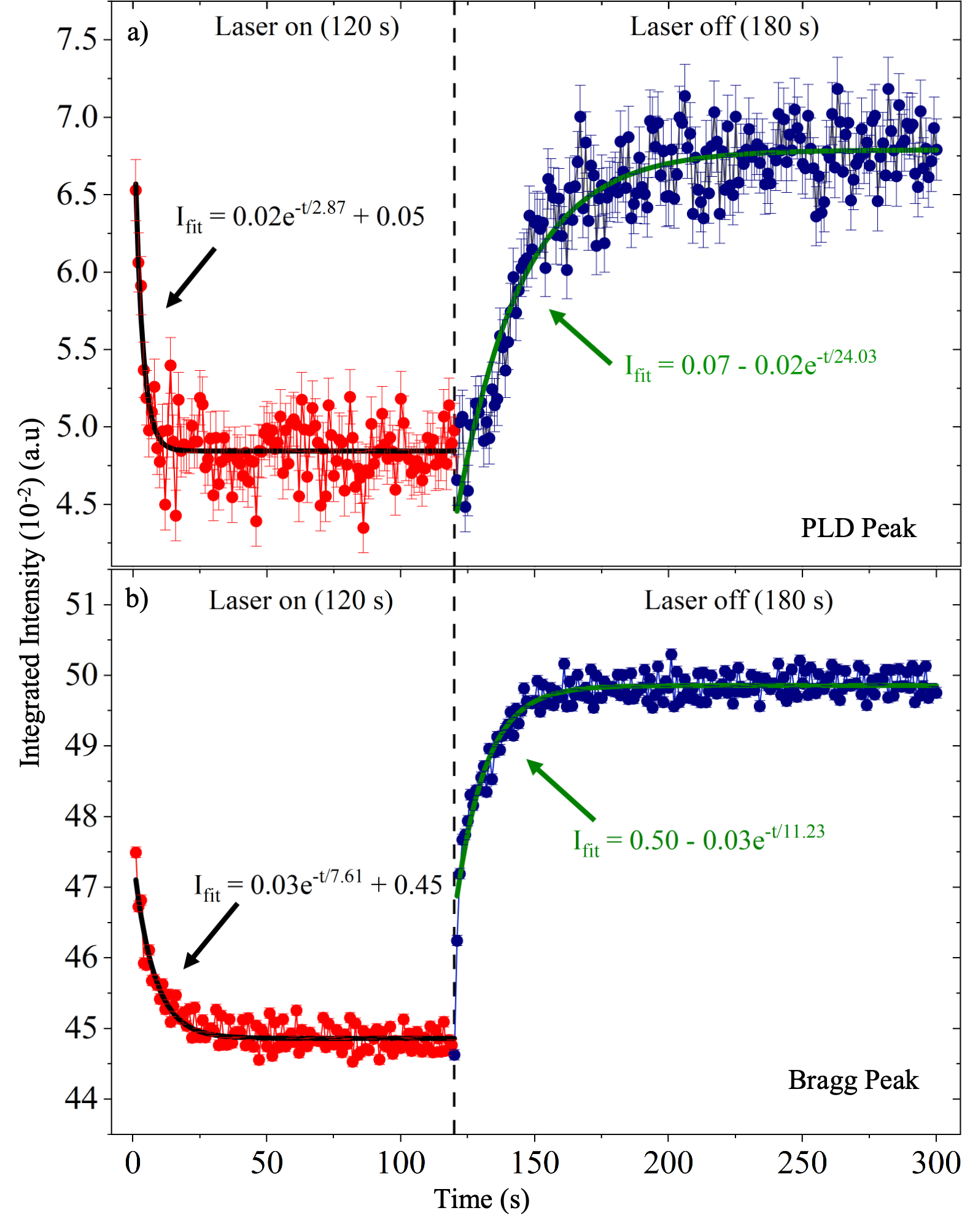}
\caption{\label{diagram} (a) A plot of the integrated intensity of the PLD and its evolution over the 300~s period. In the Laser-on phase, the intensity of the PLD peak at $\left(\frac{\bar{3}}{2}, \frac{\bar{3}}{2}, \frac{3}{2}\right)$ does not vanish with intense heating, but drops within 5~s and follows a diffusion time-constant of about 3~s. In the Laser-off phase, heat dissipation and restoration of the PLD is much slower with a time-constant of 24~s, twice as long as the Bragg peak. The intensity recovers within 30 to 40 seconds. (b) A plot of the integrated intensity and its evolution over the 300~s period. In the first 120~s, the intensity of the Bragg peak at $(\bar{2}, \bar{2}, 0)$ drops within 20~s and follows a diffusion time-constant of about 6~s. In the Laser-off phase, heat dissipation and restoration of the Bragg peak is a bit slower with a characteristic relaxation of 11~s. The lattice thermalizes quickly. The error bars are the propagated error from the neutron measurement.}
\label{fig:5vs1}
\end{figure}

The integrated intensities of the ($\frac{\bar3}{2}, \frac{\bar3}{2}, \frac{3}{2}$) superlattice and the (${\bar{2},\bar{2},0}$) fundamental Bragg peak as a function of time over the 300 s interval are shown in Fig.~\ref{fig:5vs1}, separated between the laser on and laser off modalities. During the Laser On phase, the superlattice intensity drops rapidly within the first few seconds and fluctuates near the background level thereafter, due to loss of PLD order. This behavior is described by an exponential decay, yielding a characteristic suppression time of approximately 2.9 s. At 120 s, the Laser Off phase begins, and the superlattice intensity gradually recovers, eventually saturating with a characteristic timescale of approximately 24 s. The recovery occurs over 30-40 s, significantly slower than the suppression during laser heating. This behavior is consistent with diffusive thermal relaxation and can be captured using a lumped‑capacitance heat‑transfer model, $\tau = L C_v(T)/G$, from which an interface thermal conductance of $G \approx 30$ W/m$^2$K is obtained.

\begin{figure}[H]
\centering
\includegraphics[width=4.6 in]{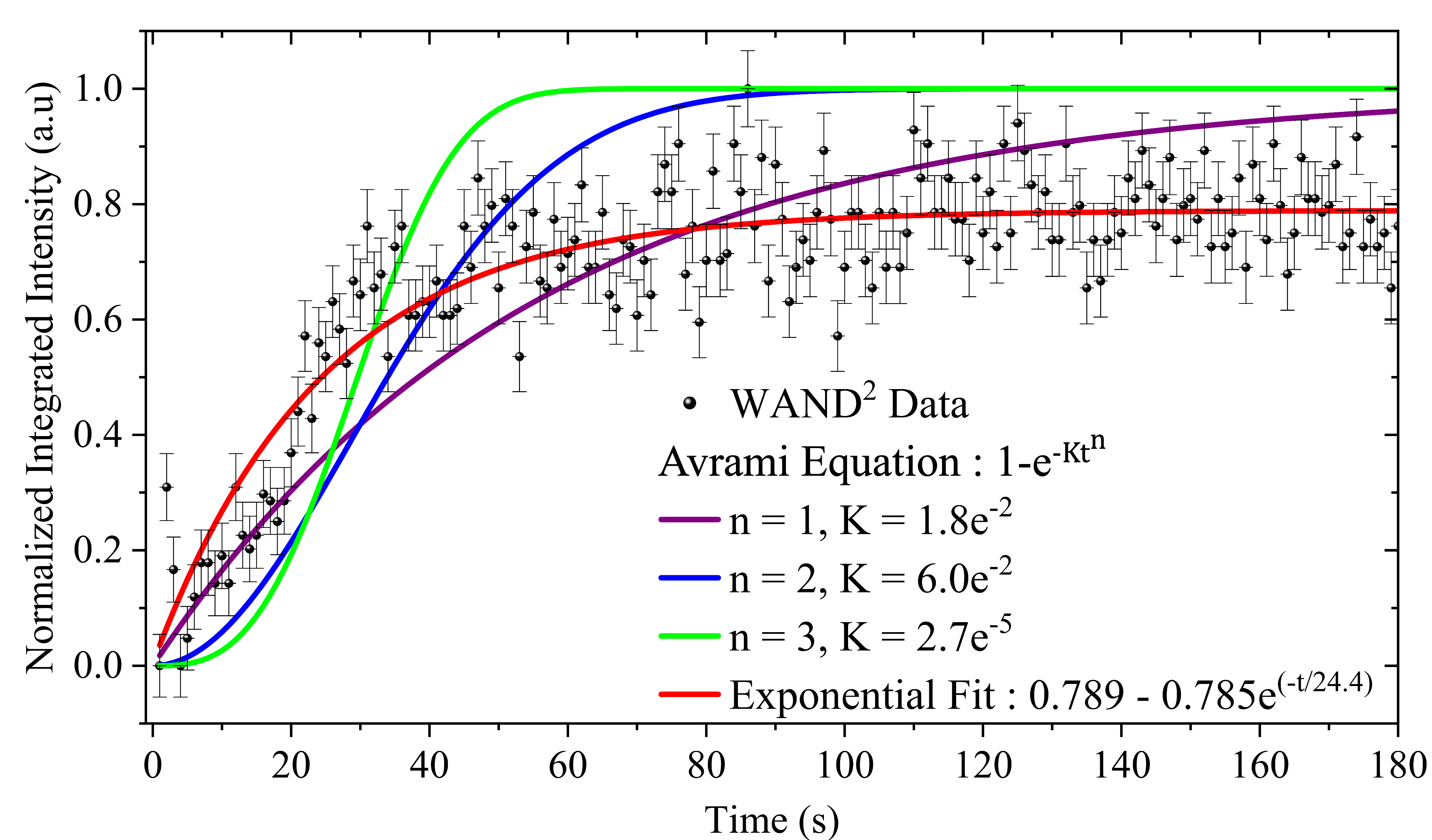}
\caption{\label{diagram} A plot of the normalized intensity in the laser off phase. The data is fit using an exponential fit as shown in the main text with a rate that is 0.04 (red line). Also compared are the Avrami equation fits for n = 1, 2 and 3. The best fit is for n=1.}
\label{fig:Avrami}
\end{figure}

Figure~\ref{fig:Avrami} is a plot of the same data shown in ~\ref{fig:Figure5} but with the intensity normalized in the laser off phase. The normalized data were fit using the Avrami equation, I(t) = 1-e$^{-Kt^{n}}$. This relation captures both the nucleation and growth rate of the CDW recovery process. The exponent, n=1, indicates a first order kinetic transformation and K = 0.04 indicates a slow recovery. The Avrami's relation with n=1 is the same as the exponential fit which corresponds to exponential kinetics. This is the simplest growth model which suggests that all nucleation sites are activated at the beginning of the laser off phase.

\begin{figure}[H]
\centering
\includegraphics[width=4.5 in]{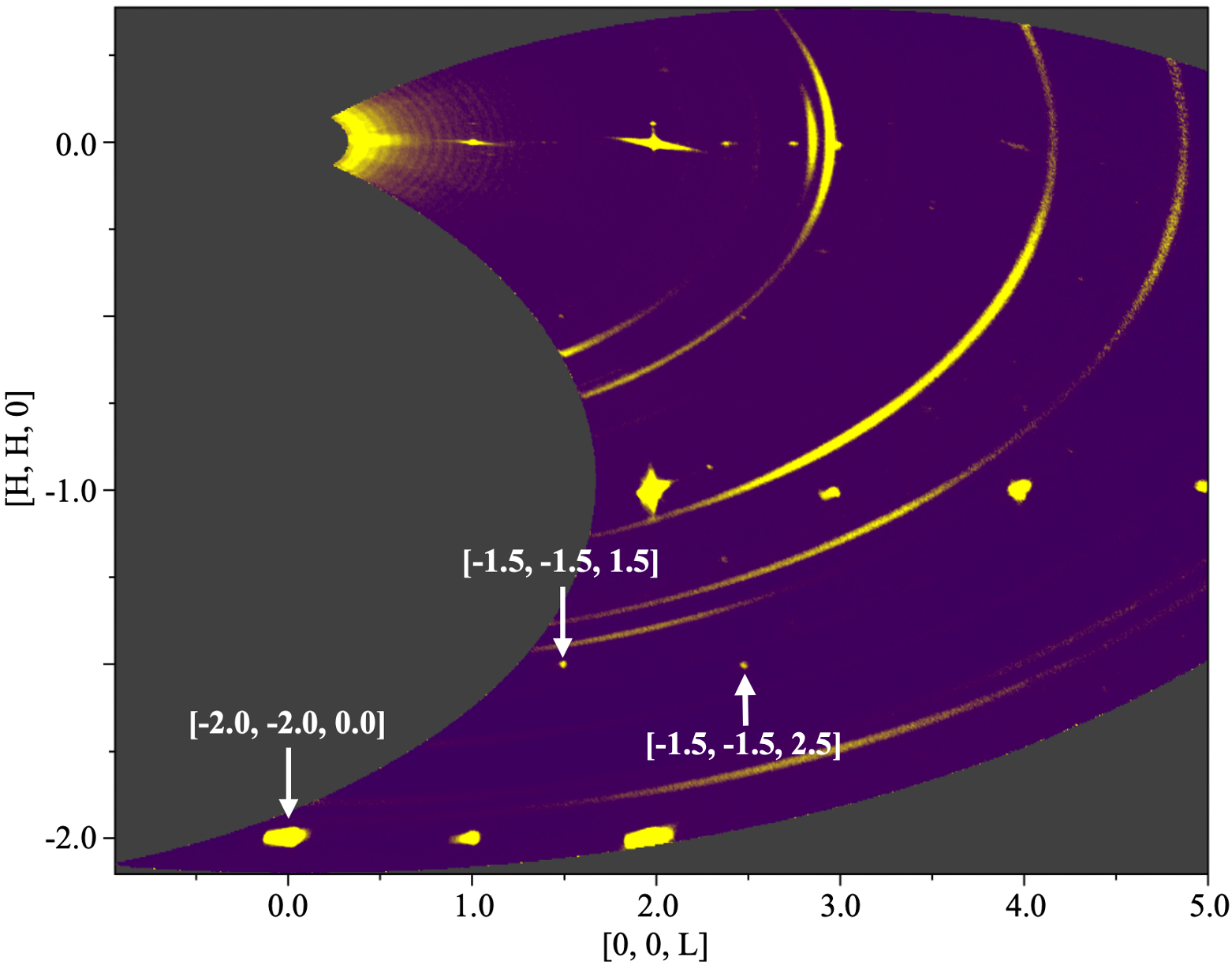}
\caption{\label{diagram} A figure of the complete scan of the TiSe${_2}$ sample to identify the CDW peaks before the laser heating.}
\label{fig:supp_bragg_peaks}
\end{figure}

Figure~\ref{fig:supp_bragg_peaks} presents a complete scan of the TiSe$_2$ sample prior to laser heating. The main Bragg peaks are clearly visible, along with the distinct superlattice peaks at [-1.5, -1.5, 1.5] and [-1.5, -1.5, 2.5]. Given the higher intensity of the $(\frac{\bar3}{2}$, $\frac{\bar3}{2}$, $\frac{3}{2})$ peak, this experiment primarily focuses on the behavior of this peak. 

\begin{figure}[H]
\centering
\includegraphics[width=4.5 in]{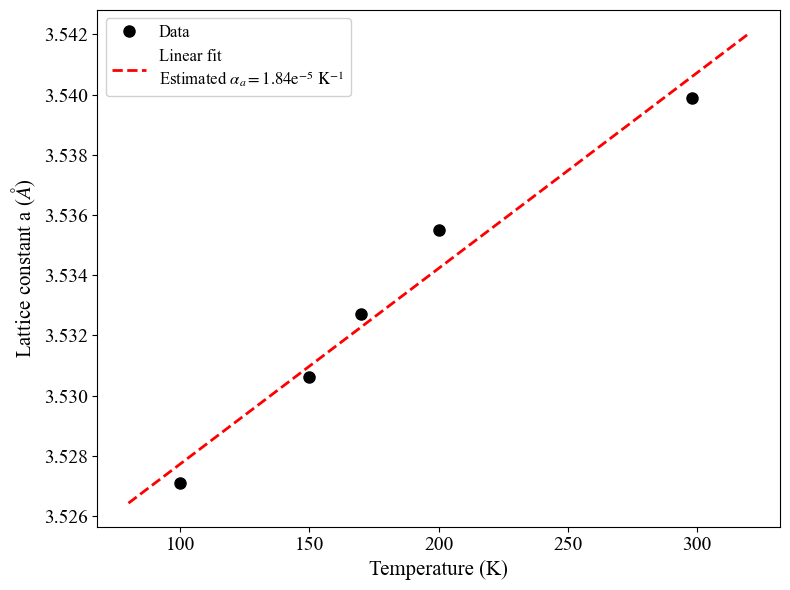}
\caption{\label{sc_xrd} The plot shows the temperature dependence of 1T-TiSe$_2$ single crystal, determined from single-crystal XRD measurements. The thermal expansion coefficient perpendicular to the unique axis ($\alpha$$_a$) was extracted from the slope of the linear fit.}
\label{fig:sc_xrd}
\end{figure}

\begin{figure}[H]
\centering
\includegraphics[width=7in]{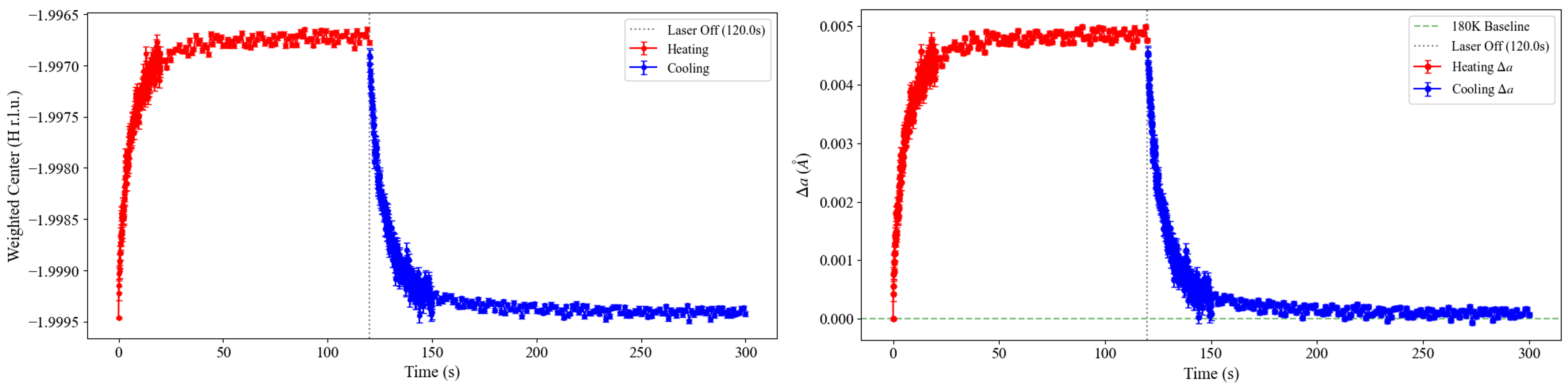}
\caption{\label{diagram} (a) Temporal shift of the Bragg peak weighted center along the H direction during heating and cooling cycles. (b) Thermal strain ($\Delta$a) of the sample as a function of time, derived from the weighted center data.}
\label{fig:weighted_center}
\end{figure}

The supplementary figures, Figure~\ref{fig:sc_xrd} and Figure~\ref{fig:weighted_center} shows the methodology used to estimate the temperature evolution of the 1T-TiSe$_2$ single crystal during laser excitation. Figure~\ref{fig:sc_xrd} establishes the equilibrium temperature dependence of the lattice constant a using single crystal XRD data, from which the thermal expansion coefficient ($\alpha_a = 1.84 \times 10^{-5}\text{ K}^{-1}$) is extracted. Figure~\ref{fig:weighted_center}(a) displays the temporal shift of the Bragg peak's weighted center in reciprocal space, capturing the rapid lattice expansion during the heating phase and the subsequent structural relaxation after the laser is turned off at 120 seconds. These reciprocal space shifts are used to determine the macroscopic dimensional changes of the lattice, which are translated into the material's real-space thermal strain ($\Delta$a) as a function of time in Figure~\ref{fig:weighted_center}(b). Finally, this time-resolved thermal strain is mathematically converted into the estimated sample temperature profile presented in Figure 4 of the main text by applying the previously calibrated $\alpha_a$.

\begin{figure}[H]
\centering
\includegraphics[width=7in]{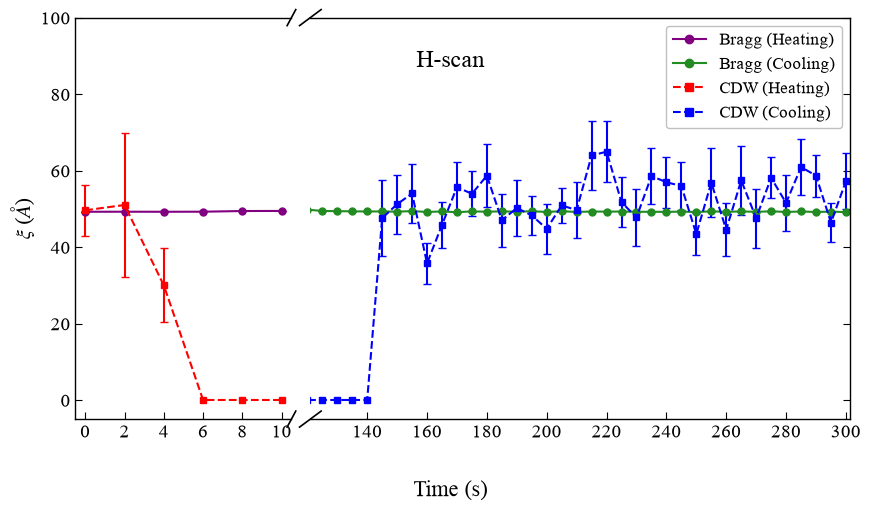}
\caption{\label{diagram} Time evolution of the correlation lengths ($\xi$) for the Bragg peak and CDW peak during the heating and cooling cycles along the H direction }
\label{fig:braggvsCDW_correlation}
\end{figure}

Figure~\ref{fig:braggvsCDW_correlation} compares the $\xi$ of the Bragg and CDW peaks during the heating and cooling cycles along the H direction. Throughout the measurement, the Bragg peak correlation length remains highly stable at approximately 50 \text{\AA}. In contrast, the CDW peak experiences a sharp collapse within 5 seconds during heating, reemerges after 30 seconds during cooling, and fluctuates around 50 \text{\AA}.

\bibliography{bibliography}